\documentclass[aps,prd,a4paper,twocolumn,amsmath,showpacs,superscriptaddress,nofootinbib,preprintnumbers]{revtex4-1}
\pdfoutput=1
\usepackage{amssymb,amsmath,latexsym,mathrsfs}
\usepackage[sort&compress]{natbib}
\usepackage{graphicx,subfigure}
\usepackage{epsfig}
\usepackage{varioref,xr-hyper}
\usepackage{color}
\usepackage{multirow}
\usepackage{array}
\usepackage{hyperref}
\usepackage{wasysym}
\usepackage{color}
\usepackage{float}
\usepackage{xcolor}
\usepackage[utf8]{inputenc}
\usepackage[T1]{fontenc}

\newcommand{\be}{\begin{equation}}
\newcommand{\ee}{\end{equation}}
\newcommand{\bea}{\begin{eqnarray}}
\newcommand{\eea}{\end{eqnarray}}

\begin{document}

\title{Fitting string inflation to real cosmological data: the Fibre Inflation case}

\author{Michele Cicoli}
\email{michele.cicoli@unibo.it}
\affiliation{Dipartimento di Fisica e Astronomia, Universit\'a di Bologna,
via Irnerio 46, 40126 Bologna, Italy and \\ INFN, Sezione di Bologna, viale Berti Pichat 6/2, 40127 Bologna, Italy }

\author{Eleonora Di Valentino}
\email{eleonora.divalentino@manchester.ac.uk}
\affiliation{Jodrell Bank Center for Astrophysics, School of Physics and Astronomy,  University of  Manchester, Oxford Road, Manchester, M13 9PL, United Kingdom}

\begin{abstract}
In this paper we show how the string landscape can be constrained using observational data. We illustrate this idea by focusing on Fibre Inflation which is a promising class of string inflationary models in type IIB flux compactifications. We determine the values of the microscopic flux-dependent parameters which yield the best fit to the most recent cosmological datasets.
\end{abstract}

\maketitle

\section{Introduction}

String theory is often said to be decoupled from experiments. However, similarly to quantum field theory, string theory is a \textit{framework} rather than a \textit{model}. Therefore it is more sensible to talk about testing a particular model built following the rules of string theory, rather than talking about testing string theory per se. 

Continuing the analogy with QFT, it is however true that features like the existence of antiparticles are common to all models which can be built in QFT. From the string theory point of view, the generic prediction is the existence of extended fundamental objects. However, due to the technical difficulty to perform experiments at energies close to the string scale, this ubiquitous prediction of string theory is presently untestable. Nonetheless we can build 4D string models and try to confront them with observations, as it is done in standard QFT with models like the Standard Model or generalisations thereof.

4D string models are characterised by interesting correlations between different observables which originate from the underlying UV consistency of the theory. These correlations can be used to compare the predictions of each 4D string model to observational data in a very efficient way, resulting in the possibility to rule out very large portions of the string landscape. 

In this paper we illustrate this idea by focusing on a class of string inflationary models called Fibre Inflation. This class of models is particularly promising since it features a landscape of examples within the framework of type IIB Large Volume flux compactifications. Each Fibre Inflation model is characterised by a different underlying choice of discrete microscopic parameters like bulk background 3-form fluxes and gauge 2-form fluxes on D-branes. 
Moreover, these models are ready to be compared with observational data since they include inflation with moduli stabilisation, consistent Calabi-Yau constructions with chiral matter and a detailed understanding of the reheating process.

In this paper we confront therefore Fibre Inflation with the most recent cosmological observations including data from Planck, local measurements of the Hubble constant, Baryon Acoustic Oscillation, the Dark Energy Survey and CMB lensing. In doing so, we find the model in the Fibre Inflation landscape which gives the best fit to these cosmological datasets. Considering Planck 2018 temperature and polarisation data only, the bounds for the main cosmological observables at $68\%$ CL become $n_s = 0.9696^{+0.0010}_{-0.0026}$ and $r = 0.00731^{+0.00026}_{-0.00072}$ together with a number of effective relativistic species $N_{\rm eff} = 3.062^{+0.004}_{-0.015}$. These predictions, in turn, constrain the microscopic flux-dependent parameters.

This paper is organised as follows. In Sec.~\ref{FibreReview} we briefly review the main features of Fibre Inflation models while in Sec.~\ref{Methodology} we describe the methodology of our analysis. Our results are presented in Sec. \ref{Results} and in Sec. \ref{MicroscopicBounds} they are translated into bounds on the microscopic flux-dependent parameters. Finally in Sec.~\ref{Conclusions} we discuss our results and present our conclusions.

\section{Fibre Inflation models: a brief overview}
\label{FibreReview}

Fibre Inflation (FI) is a class of string inflationary models built within the framework of type IIB Large Volume Scenarios \cite{Balasubramanian:2005zx, Cicoli:2008va}. Its name comes from the fact that the inflaton is a K\"ahler modulus which controls the size of a K3 or T$^4$ divisor fibred over a $\mathbb{P}^1$ base. 

In the original model the inflationary potential is generated by a combination of 1-loop open string corrections~\cite{Cicoli:2008gp}. However subsequent examples of FI models have been constructed by exploiting different combinations of perturbative corrections to the effective action~\cite{Broy:2015zba, Cicoli:2016chb}. This class of models originates very naturally due to the presence of an effective shift symmetry which protects the flatness of the inflationary potential~\cite{Burgess:2016owb, Burgess:2014tja}. Moreover FI models are not just at the level of a string-inspired 4D supergravity since in~\cite{Cicoli:2011it, Cicoli:2016xae, Cicoli:2017axo} they have been embedded into globally consistent Calabi-Yau orientifolds with a chiral brane set-up and moduli stabilisation.

FI models have a behaviour similar to Starobinsky inflation~\cite{Starobinsky:1980te} and supergravity $\alpha$-attractors~\cite{Kallosh:2013maa,Kallosh:2017wku}. In fact, their potential features a trans-Planckian plateau which steepens at large inflaton field values due to higher derivative or loop corrections that can be responsible for a CMB power loss at large angular scales~\cite{Cicoli:2013oba, Pedro:2013pba, Cicoli:2014bja}. The inflaton field range is constrained also by geometrical bounds~\cite{Cicoli:2018tcq} but in generic FI models it is around $5$ in Planck units. This inflaton excursion yields primordial gravity waves at the edge of detectability since the tensor-to-scalar ratio is of order $0.007 \lesssim r \lesssim 0.01$.

The exact prediction for the amplitude of the primordial tensor modes needs a proper understanding of the post-inflationary evolution of these models. Relatively recently, ref.~\cite{Antusch:2017flz} found that preheating effects can be neglected in FI models and ref.~\cite{Cabella:2017zsa, Cicoli:2018cgu} performed a detailed analysis of perturbative reheating finding that the inflaton decay can produce a thermal bath with an initial temperature which is below the maximal one derived from requiring that finite-temperature corrections to the inflationary potential do not induce a decompactification limit~\cite{Anguelova:2009ht}. Moreover the inflaton decay tends to produce, on top of ordinary particles, also hidden sector degrees of freedom like ultra-light bulk axions~\cite{Cicoli:2018cgu} which behave as extra dark radiation parameterised by $\Delta N_{\rm eff}$. 

From the phenomenological point of view, it is very interesting to notice that values of $\Delta N_{\rm eff}\simeq 0$ correlate with values of $r\simeq 0.007$ and no CMB power loss at large angular scales, while values of $\Delta N_{\rm eff}\simeq 0.5$ correlate with values of $r\simeq 0.01$ and a low-$\ell$ CMB power loss. From the theoretical point of view, different values of $\Delta N_{\rm eff}$ and $r$ correspond to different choices for the underlying UV parameters. In this paper we will confront FI models with cosmological observations and see which values of $\Delta N_{\rm eff}$ and $r$ give the best fit to actual data. This will allow us to constrain the values of the stringy parameters and to judge the naturalness of these models from the theoretical point of view.

\subsection{Inflationary potential and observables}

\begin{figure}[!htbp]
\centering
\includegraphics[scale = 0.7]{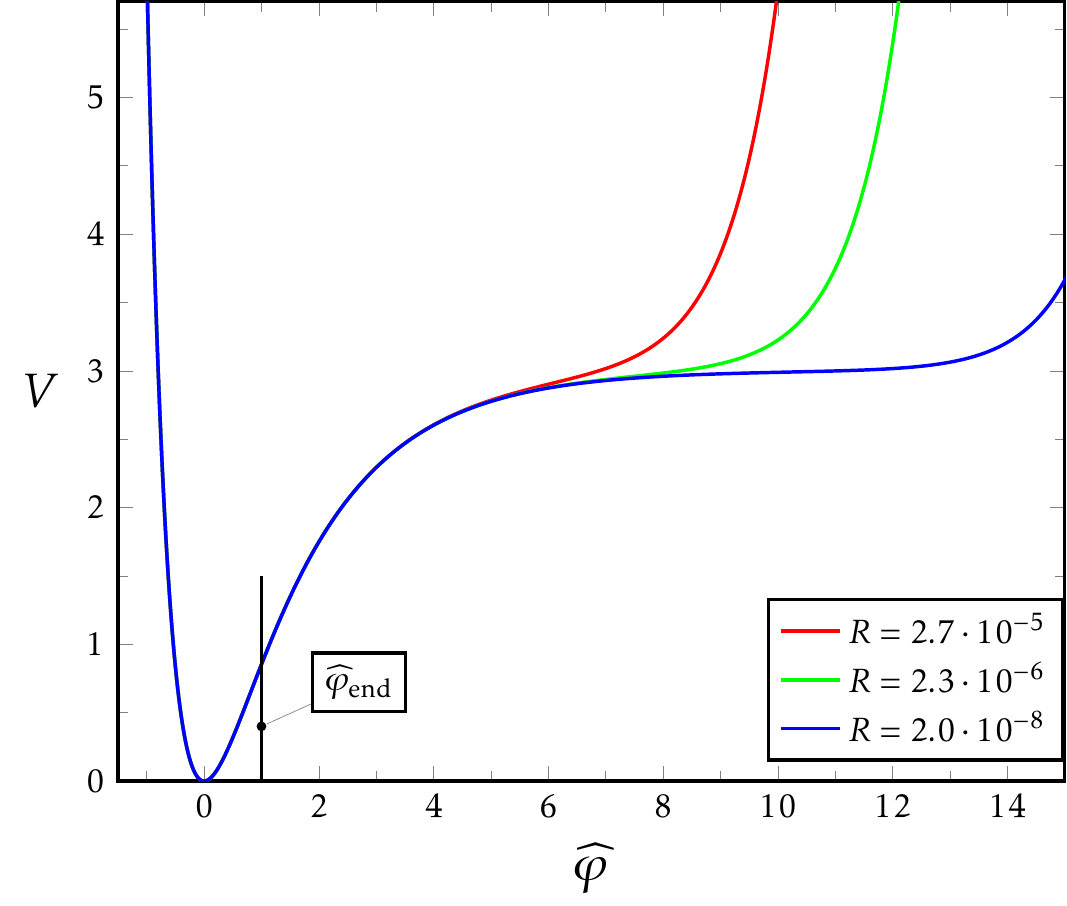}
\caption{Inflationary potential in Planck units for different values of $R$ setting $V_0 = 1$. The plot shows also the end point of inflation.}
\label{Fig1}
\end{figure}

All FI models feature a qualitatively similar shape of the inflationary potential. Without loss of generality, we therefore focus on the potential of the original model~\cite{Cicoli:2008gp} which looks like (where $M_p$ is the reduced Planck mass $M_p = \frac{1}{\sqrt{8\pi G}}\simeq 2.4\cdot 10^{18}$ GeV):
\be 
V(\phi) =V_0 M_p^4 \,U(\phi)
\ee
with:
\be
 U(\phi) = 3-4 \,e^{-\frac{\phi}{M_p\sqrt{3}} }  +  e^{-\frac{4\phi}{M_p\sqrt{3}} }+ R\left(e^{\frac{2\phi}{M_p\sqrt{3}} } - 1\right)
\label{Inflationpot}
\ee
where $V_0$ and $R$ are two independent parameters which depend on different combinations of the microscopic parameters. In the regime where the effective field theory is under control both $V_0\ll 1$ and $R\ll 1$. Fig.~\ref{Fig1} shows the potential (\ref{Inflationpot}) for different values of $R$. 

The expression for the number of efoldings $N_e$ as a function of the point of horizon exit in field space $\phi_*$ cannot be solved analytically. Following~\cite{Cicoli:2018cgu}, we therefore consider a simplified case where an approximated analytical solution can be provided. 

The scalar spectral index takes the form: 
\be
n_s (N_e, R) = 1- \frac89\,C  - \frac{16}{9}\,C^2\,\left(1+D\right)\,\left(1+\frac{R}{2}\, e^{\frac{\sqrt{3}\phi_*}{M_p}}\right)^2 \,,
\label{nsrnew}
\ee
where:
\bea
D &=& D (\phi_*,R) = \sum_{n=1}^\infty (n+1) R^n e^{n\frac{\sqrt{3}\phi_*}{M_p}} \nonumber \\
C &=& C (\phi_*, R) = e^{-\frac{\phi_*}{M_p\sqrt{3}} } -R \,e^{\frac{2\phi_*}{M_p\sqrt{3}}} \nonumber \\
\phi_* &=& \phi_* (N_e, R) = \sqrt{3} M_p \ln\left\{f +\frac43\ln f \right.  \nonumber \\
 &- & \frac13 \sum_{n=1}^\infty (-1)^n 
\left[\frac{3}{3n+1} \left(f^{1+3n}-e^{\frac{1}{\sqrt{3}}+n\sqrt{3}}\right)\right. \nonumber \\
&-& \left. \left.\frac{2}{n}  \left( f^{3n}-e^{n\sqrt{3}} \right)\right]   \left(\frac{R}{2}\right)^n\right\} \nonumber \\
f &=& f(N_e) = \frac49\,N_e + e^{\frac{1}{\sqrt{3}}} -\frac{4}{3\sqrt{3}} \nonumber \,. \label{phistNeR}
\eea
The tensor-to-scalar ratio is given by:
\bea
r(N_e,R) &=& 6\left(n_s  -1\right)^2\,\left(1+D\right)\left(1+\frac{R}{2}\, e^{\frac{\sqrt{3}\phi_*}{M_p}}\right)^2 ,
\label{nsr2}
\eea
while the scalar power spectrum reads:
\be
P(k) = A_s \left(\frac{k}{k_*}\right)^{n_s-1}\,,
\ee
with $k_*=0.05\,{\rm Mpc}^{-1}$ and:
\be
A_s = A_s(N_e,R,V_0) = \frac{V(\phi=\phi_*)}{24 \pi^2 M_p^4 \epsilon}\, \,,
\label{As}
\ee
where:
\be
\epsilon(\phi_*,R) = \frac{8}{27}\,C^2\left(1+D\right)\left(1+\frac{R}{2}\, e^{\frac{\sqrt{3}\phi_*}{M_p}}\right)^2\,.
\ee

\begin{figure}[!htbp]
\centering
\includegraphics[scale = 0.6]{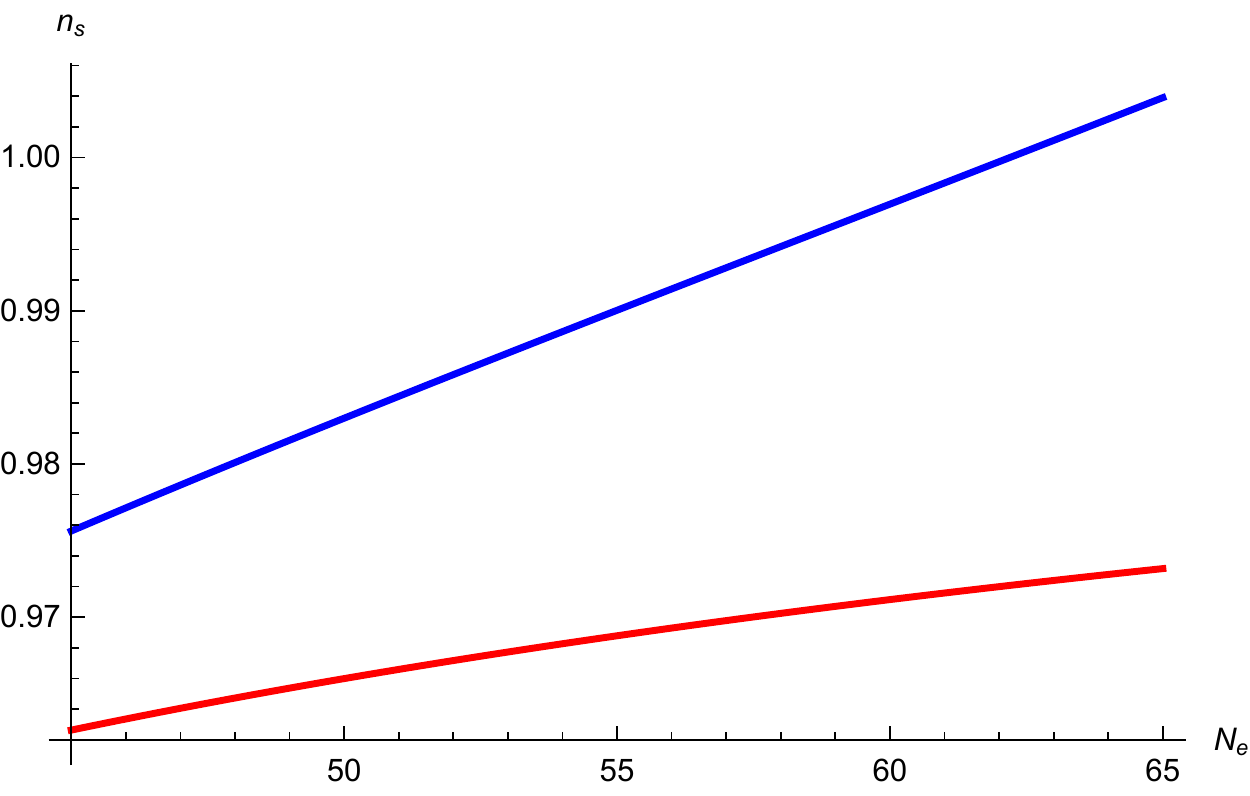}
\includegraphics[scale = 0.6]{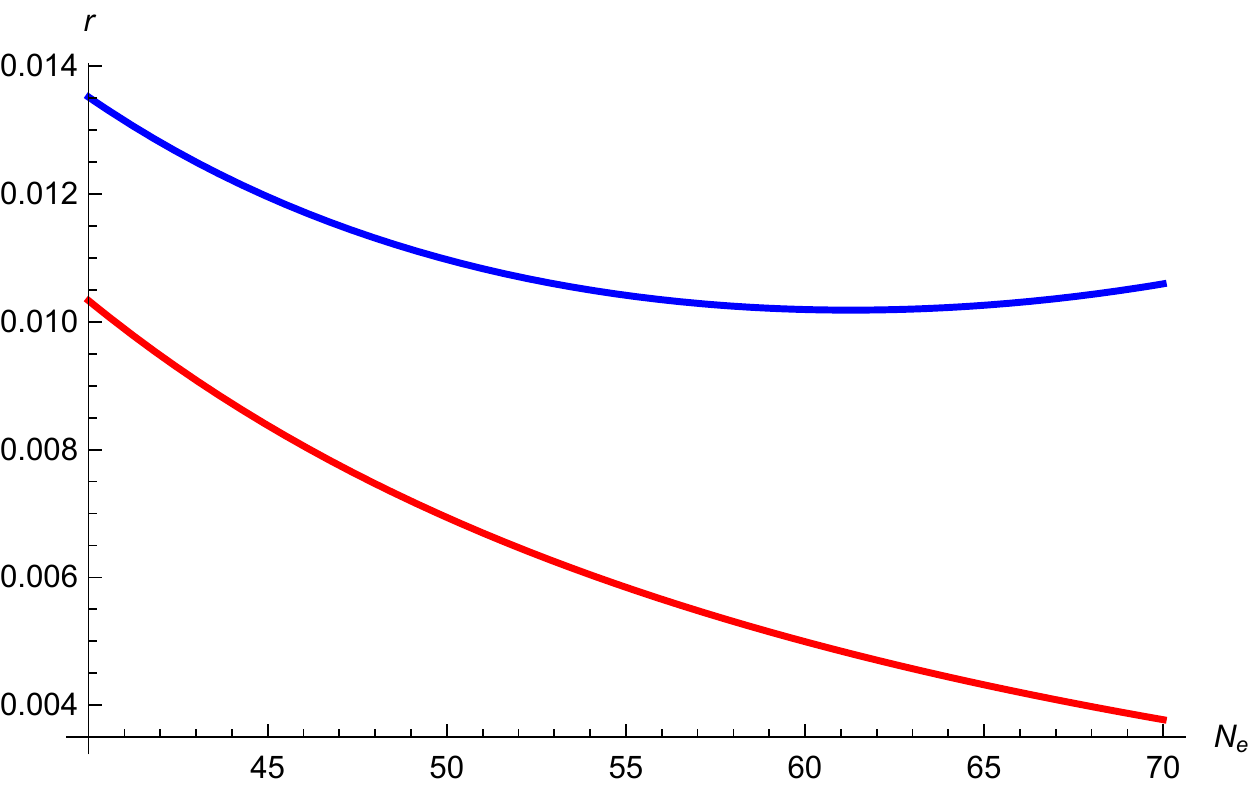}
\caption{$n_s$ and $r$ as functions of $N_e$ for $R= 2.7\cdot 10^{-5}$ (blue lines) and $R=0$ (red lines).}
\label{Fig1b}
\end{figure}

Fig.~\ref{Fig1b} shows $n_s$ and $r$ as functions of $N_e$ for $R= 2.7\cdot 10^{-5}$ (blue lines) and $R=0$ (red lines). The red line represents the case where the positive exponential in (\ref{Inflationpot}) is negligible throughout the whole inflationary dynamics. Notice that, for the same value of $N_e$, the red line gives smaller $n_s$ and $r$. In particular, for the red line, $n_s$ cannot be larger than about $0.97$ for $N_e\lesssim 65$. 

\begin{figure}[!htbp]
\centering
\includegraphics[scale = 0.7]{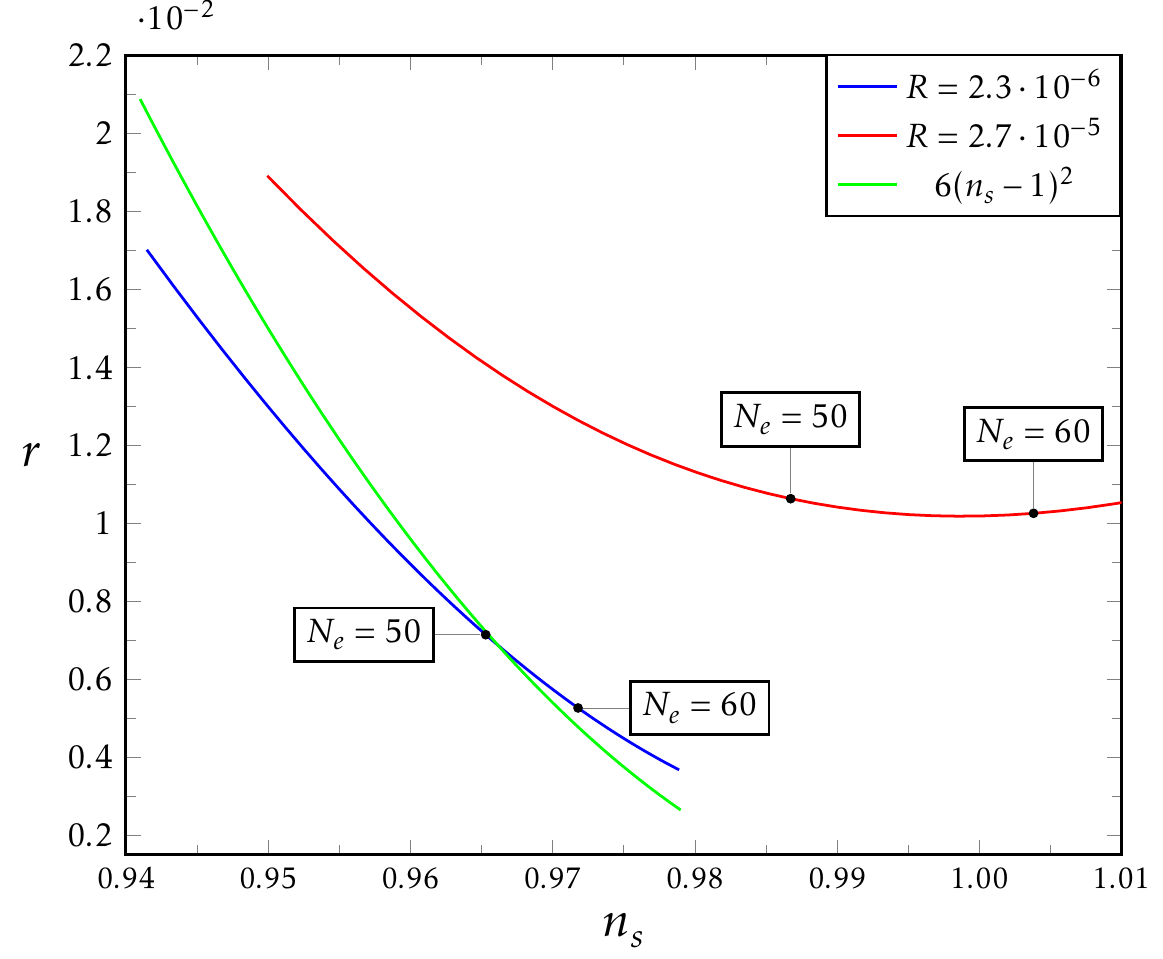}
\caption{$r$ as a function of $n_s$ for two different values of $R$ (red and blue lines). The green curve represents the relation $r = 6(n_s -1)^2$.}
\label{Fig3}
\end{figure}

Fig.~\ref{Fig3} shows instead $r$ versus $n_s$ for two different values of $R$. The green curve represents the relation $r=6(n_s-1)^2$ which is a good approximation for $R=0$. Interestingly, values of $R$ of order $R=2.3\cdot 10^{-6}$ agree with the green curve rather well while for $R=2.7\cdot 10^{-5}$ already the relation $r=6(n_s-1)^2$ is violated.

\subsection{Number of efoldings, reheating and dark radiation}
\label{SecB}

\begin{figure}[h!]
\begin{center}
	\includegraphics[width=0.97\linewidth]{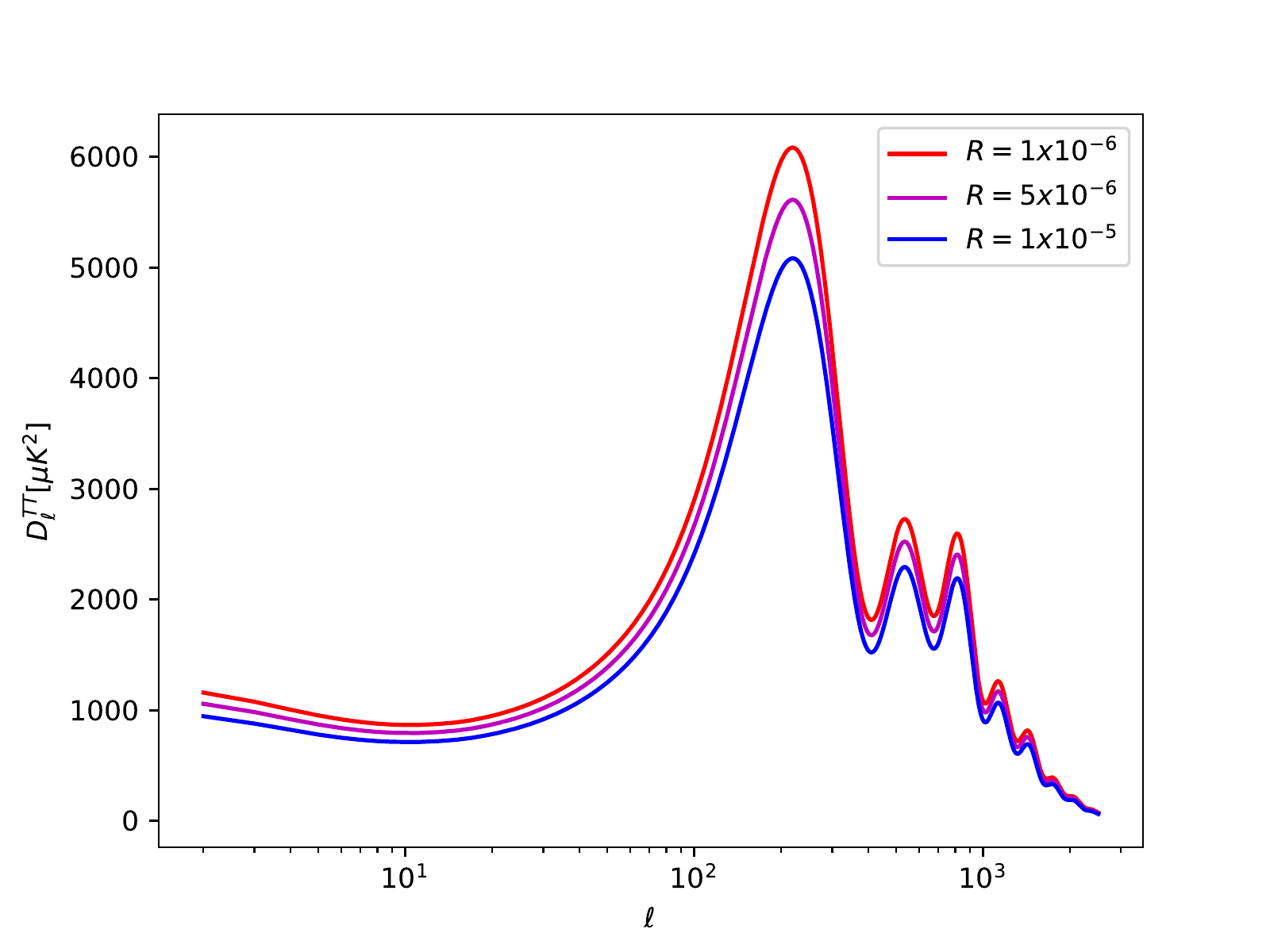}
	\includegraphics[width=0.97\linewidth]{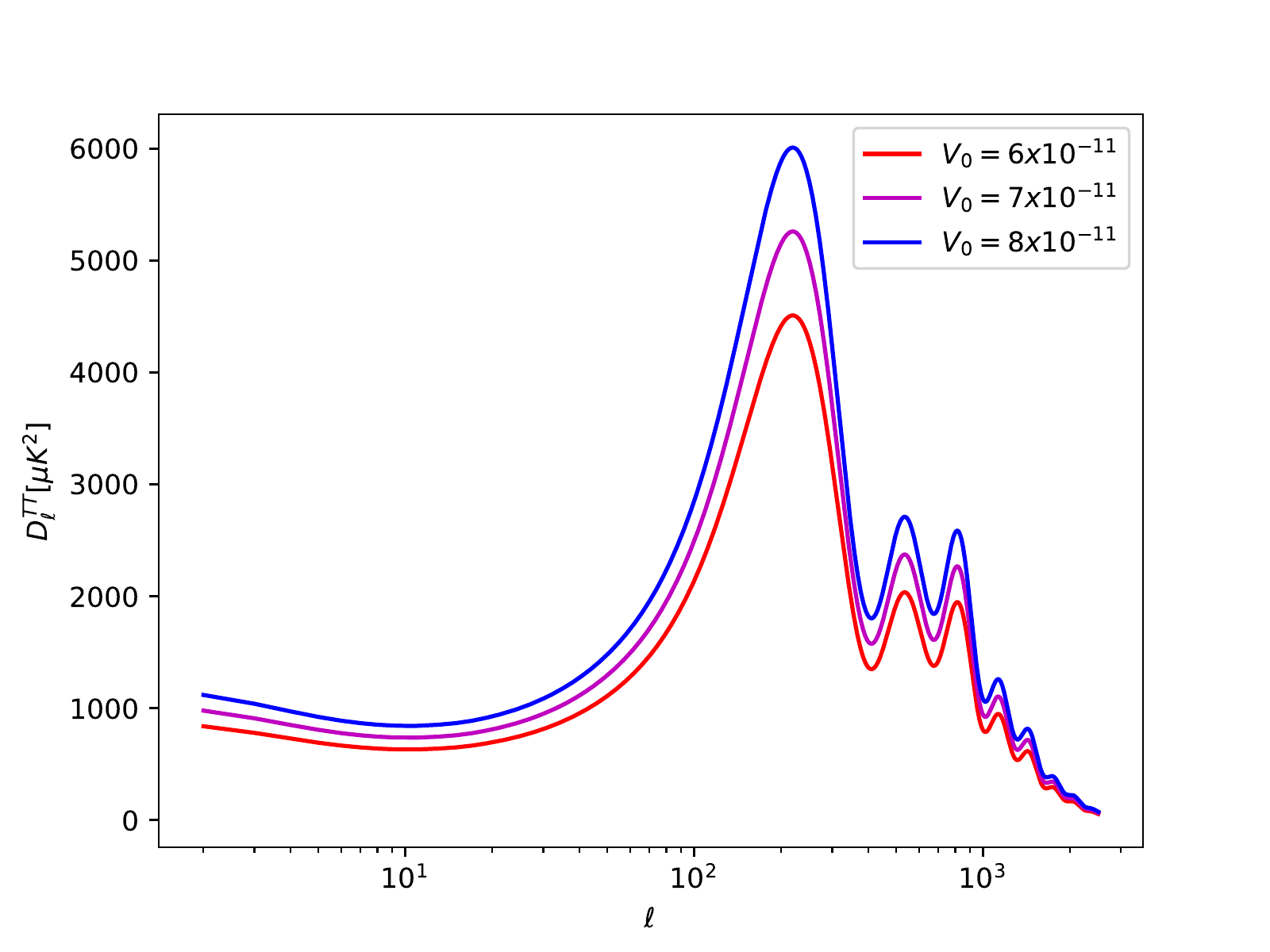}
	\includegraphics[width=0.97\linewidth]{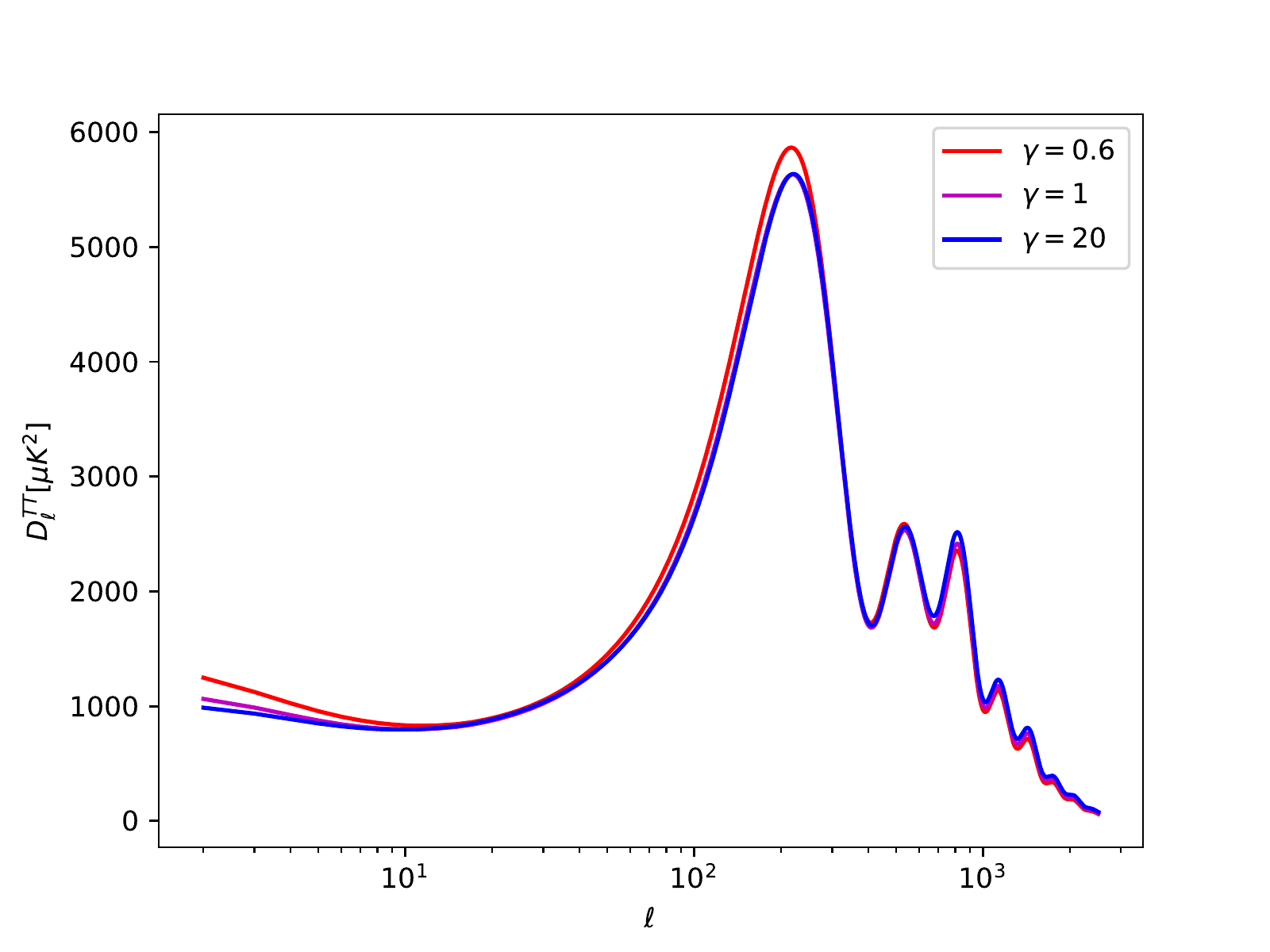}
	\caption{Theoretical variation of the temperature power spectrum obtained by varying the microscopic parameters of FI models: $R$, $V_0$ and $\gamma$. $R$ (top panel) affects the temperature power spectrum increasing the amplitude of the peaks when decreasing. $V_0$ (middle panel) does exactly the opposite, increasing the amplitude of the peaks when increasing. Finally, $\gamma$ (bottom panel) suppresses the amplitude of the low multipole range when increasing.}
	\label{fig:TT}
\end{center}
\end{figure}

After the end of inflation the inflaton oscillates around the minimum and behaves as non-relativistic matter. Hence reheating is characterised by an equation of state $p = w_{\rm rh}\rho$ with $w_{\rm rh}=0$. Moreover the inflationary energy scale $M_{\rm inf}^4 = V(\phi=\phi_*)$ turns out to be around the GUT scale. Thus $N_e$ can be written in terms of the reheating temperature $T_{\rm rh}$ as:
\be
N_e \simeq 58 -\frac13\ln\left(\frac{M_p}{T_{\rm rh}}\right).
\label{Ne1}
\ee
In turn, the reheating temperature can be written as:
\be
T_{\rm rh} = 3\,\gamma\cdot 10^{10}\,{\rm GeV} \,,
\label{TrhFIfin}
\ee
where $\gamma$ is a parameter independent from $V_0$ and $R$ which controls the branching ratios for the inflaton decay into different visible and hidden sector degrees of freedom \cite{Cicoli:2018cgu}. This gives:
\be
N_e =  52+\frac13 \ln\gamma \,. 
\ee
Plugging this value in (\ref{nsrnew}), (\ref{nsr2}) and (\ref{As}), the cosmological observables become functions of the underlying parameters $\gamma$, $R$ and $V_0$: $n_s=n_s(\gamma,R)$, $r=r(\gamma,R)$ and $A_s = A_s(\gamma,R,V_0)$. In Fig.~\ref{fig:TT} we show how these parameters $\gamma$, $R$ and $V_0$ affect the CMB temperature power spectra. Therefore they can be constrained by the requirement of matching Planck data which however depend on the number of extra neutrino-like species $\Delta N_{\rm eff}$ that in this model is given by:
\be
\Delta N_{\rm eff} = \frac{0.6}{\gamma^2}\,, 
\label{NeffPred}
\ee

As studied in \cite{Cicoli:2018cgu}, the main inflaton decay channels are into Standard Model gauge bosons and hidden sector ultra-light axions. In the data analysis which we will present in Sec.~\ref{Results}, we will focus on the following three different ranges for $\gamma$:
\begin{enumerate}
\item $1 < \gamma\leq 20$ corresponds to the case where the branching ratio for the inflaton decay into hidden axions is suppressed by a non-zero gauge flux on the D7-brane stack wrapped around the inflaton divisor which realizes the Standard Model. Therefore in this case $\Delta N_{\rm eff}$ is negligibly small. The upper bound $\gamma\leq 20$ can be easily derived by combining moduli stabilisation with two requirements: ($i$) a correct matching of the observed value of the density perturbations, and ($ii$) an effective field theory which remains in the controlled regime where perturbation theory does not break down. See App.~\ref{AppA} for technical details.

\item $\gamma = 1$ corresponds to the case where the branching ratio for the inflaton decay into hidden axions is maximized due to the fact that the gauge flux on the Standard Model D7-branes is vanishing. In this case $\Delta N_{\rm eff} = 0.6$. 

\item $0<\gamma<1$ corresponds to the case where the amount of extra dark radiation gets further increased by the model-dependent presence of additional inflaton decay channels into hidden sector degrees of freedom like for example hidden gauge bosons living on D7-branes wrapped around the divisor containing the $\mathbb{P}^1$ base of the fibration. 
\end{enumerate}

Ref.~\cite{Cicoli:2018cgu} pointed out that there are two qualitatively different regimes:
\begin{itemize}
\item \textbf{Small extra dark radiation:} If $\gamma\gtrsim 2$, $\Delta N_{\rm eff}\lesssim 0.1$ which requires a spectral index centered around $n_s\simeq 0.965$ \cite{Aghanim:2018eyx}. As can be seen from Fig. \ref{Fig1b} and \ref{Fig3}, this can be achieved if $R\ll  2.7\cdot 10^{-5}$ (notice that from the microscopic point of view, no fine-tuning is involved to get such a small value of $R$). In this case horizon exit takes place in the plateau region where $r\simeq 0.007$. An explicit example presented in \cite{Cicoli:2018cgu} is: 
\be
\qquad R = 1.78\cdot 10^{-7} \quad V_0 = 7.78 \cdot 10^{-11}\quad \gamma=3.316\,, \nonumber
\ee
which gives $\phi_{\rm end} = 0.917\,M_p$ where $\epsilon(\phi_{\rm end})=1$ and $\phi_* = 5.801\,M_p$ where $N_e(\phi_*)=52$ and:
\be
\qquad \Delta N_{\rm eff} \simeq 0.05 \qquad n_s=  0.965 \qquad r=0.0065\,. \nonumber
\ee

\item \textbf{Large extra dark radiation:} For $\gamma\simeq 1$, the amount of extra dark radiation is larger since $\Delta N_{\rm eff}\lesssim 0.5$. Ref. \cite{Cicoli:2018cgu} used 2015 Planck data to infer that such a large value of $\Delta N_{\rm eff}$ requires a spectral index centered around $n_s\simeq 0.99$ \cite{Ade:2015xua}. As shown in Fig. \ref{Fig1b} and \ref{Fig3}, this is possible if $R\gtrsim 2.7\cdot 10^{-5}$. In this case the potential is steeper close to horizon exit, and so $r\simeq 0.01$. An explicit example presented in \cite{Cicoli:2018cgu} is: 
\be
\qquad R = 2.76\cdot 10^{-5} \quad V_0 = 1.24 \cdot 10^{-10}\quad \gamma=1.268\,, \nonumber
\ee
which gives $\phi_{\rm end} = 0.918\,M_p$ where $\epsilon(\phi_{\rm end})=1$ and $\phi_* = 5.945\,M_p$ where $N_e(\phi_*)=52$ and:
\be
\qquad \Delta N_{\rm eff} \simeq 0.37 \qquad n_s=  0.99 \qquad r=0.01\,. \nonumber
\ee

However, more recent 2018 Planck data including \textit{lowE} polarization data~\cite{Aghanim:2018eyx} give a strong lower constraints on the optical depth $\tau$ than 2015 Planck measurements. For this reason, and thanks to the positive correlation with $N_{\rm eff}$ and $n_s$, both the extra dark radiation component $\Delta N_{\rm eff}$ and the spectral index $n_s$ will have lower mean values. In fact, as we shall see in our analysis in Sec. \ref{Results}, when $\gamma\simeq 1$, the central value of the spectral index raises to $n_s\simeq 0.973$ but not more. In turn this gives a tensor-to-scalar ratio of order $r\simeq 0.0085$ which requires values of $R$ slightly smaller than $R\gtrsim 2.7\cdot 10^{-5}$.
\end{itemize}

\section{Methodology}
\label{Methodology}

To analyse the effect of the FI scenario on the constraints on the cosmological parameters, we consider some of the most recent cosmological datasets, listed below:

\begin{itemize}

\item {\bf Planck}: we make use as a baseline of the Planck $2018$ temperature and polarization CMB angular power spectra {\it plikTTTEEE+lowl+lowE}~\cite{Aghanim:2018eyx,Aghanim:2019ame}.

\item {\bf R19}: we assume a gaussian prior on the Hubble constant $H_0$ as obtained from the SH0ES collaboration in~\cite{Riess:2019cxk}, i.e. $H_0=74.03\pm1.42$ (km/s)/Mpc at 68\% CL.

\item {\bf BAO}: we consider the Baryon Acoustic Oscillation data from the same compilation adopted in~\cite{Aghanim:2018eyx}, composed of the 6dFGS~\cite{Beutler:2011hx}, SDSS MGS~\cite{Ross:2014qpa}, and BOSS DR12~\cite{Alam:2016hwk} data.

\item {\bf DES}: we add the $3\times2$pt analysis of the first-year of the Dark Energy Survey measurements~\cite{Troxel:2017xyo, Abbott:2017wau, Krause:2017ekm}, as used in~\cite{Aghanim:2018eyx}.

\item {\bf lensing}: we use the 2018 CMB lensing reconstruction power spectrum as obtained from the CMB trispectrum analysis~\cite{Aghanim:2018oex}.

\item {\bf Pantheon}: we consider the luminosity distance measurements of $1048$ type Ia Supernovae from the Pantheon catalog~\cite{Scolnic:2017caz}.

\end{itemize}

We consider as a baseline a 7-dimensional parameter space described by: the baryon and cold dark matter energy densities $\Omega_bh^2$ and $\Omega_{c}h^2$, the ratio of the sound horizon at decoupling to the angular diameter distance to last scattering $100 \theta_{MC}$, the optical depth to reionization $\tau$, and three combinations of microscopic parameters characterising FI models: $\gamma$, $R$ and $V_0$. We impose flat uniform priors on these parameters, as showed in Table~\ref{priors}, where we distinguish three different cases, depending on the range of $\gamma$. In our analysis the amplitude and the spectral index of the primordial scalar perturbations $A_s$ and $n_s$, as well as the effective number of relativistic degrees of freedom $N_{\rm eff}$ and the tensor-to-scalar ratio $r$ are, instead, derived parameters, obtained by using eqs. (\ref{As}), (\ref{nsrnew}), (\ref{nsr2}) and (\ref{NeffPred}) respectively. In order to do this computation, we stop the series used in the code at $n=65$, after checking that the changes on the parameters for a larger number were below the numerical sensitivity of the code.

\begin{table}
\begin{center}
\begin{tabular}{cccc}
\hline
Parameter                    & prior& prior & prior\\
\hline
$\Omega_{b} h^2$ & $[0.005,0.1]$& $[0.005,0.1]$& $[0.005,0.1]$\\
$\Omega_{c} h^2$ & $[0.001,0.99]$& $[0.001,0.99]$& $[0.001,0.99]$   \\
$\tau$ & $[0.01,0.8]$& $[0.01,0.8]$& $[0.01,0.8]$   \\
$100\theta_{MC}$ & $[0.5,10]$& $[0.5,10]$& $[0.5,10]$   \\
$\gamma$ & $[0, 1]$& $=1$&$[1,20]$   \\
$R$ & $[0, 10^{-5}]$& $[0, 10^{-5}]$ & $[0, 10^{-5}]$\\
$10^{11}V_0$ & $[1,10]$& $[1,10]$& $[1,10]$\\
\hline
\end{tabular}
\end{center}
\caption{Flat priors on the cosmological parameters assumed in this work.}
\label{priors}
\end{table}

In order to extract the posterior distribution of these cosmological parameters, we use our modified version of the publicly available Monte-Carlo Markov Chain package \texttt{CosmoMC}~\cite{Lewis:2002ah}, with a convergence diagnostic based on the Gelman and Rubin statistics~\cite{Gelman:1992zz}, that implements an efficient sampling of the posterior distribution using the fast/slow parameter decorrelations \cite{Lewis:2013hha}, and includes the support for the 2018 Planck data release~\cite{Aghanim:2019ame} (see \url{http://cosmologist.info/cosmomc/}).

\section{Results}
\label{Results}

In this section we show and discuss the results for the three different ranges of $\gamma$.

\subsection{$0 < \gamma < 1$}

\begin{figure*}
\begin{center}
	\includegraphics[width=0.7\linewidth]{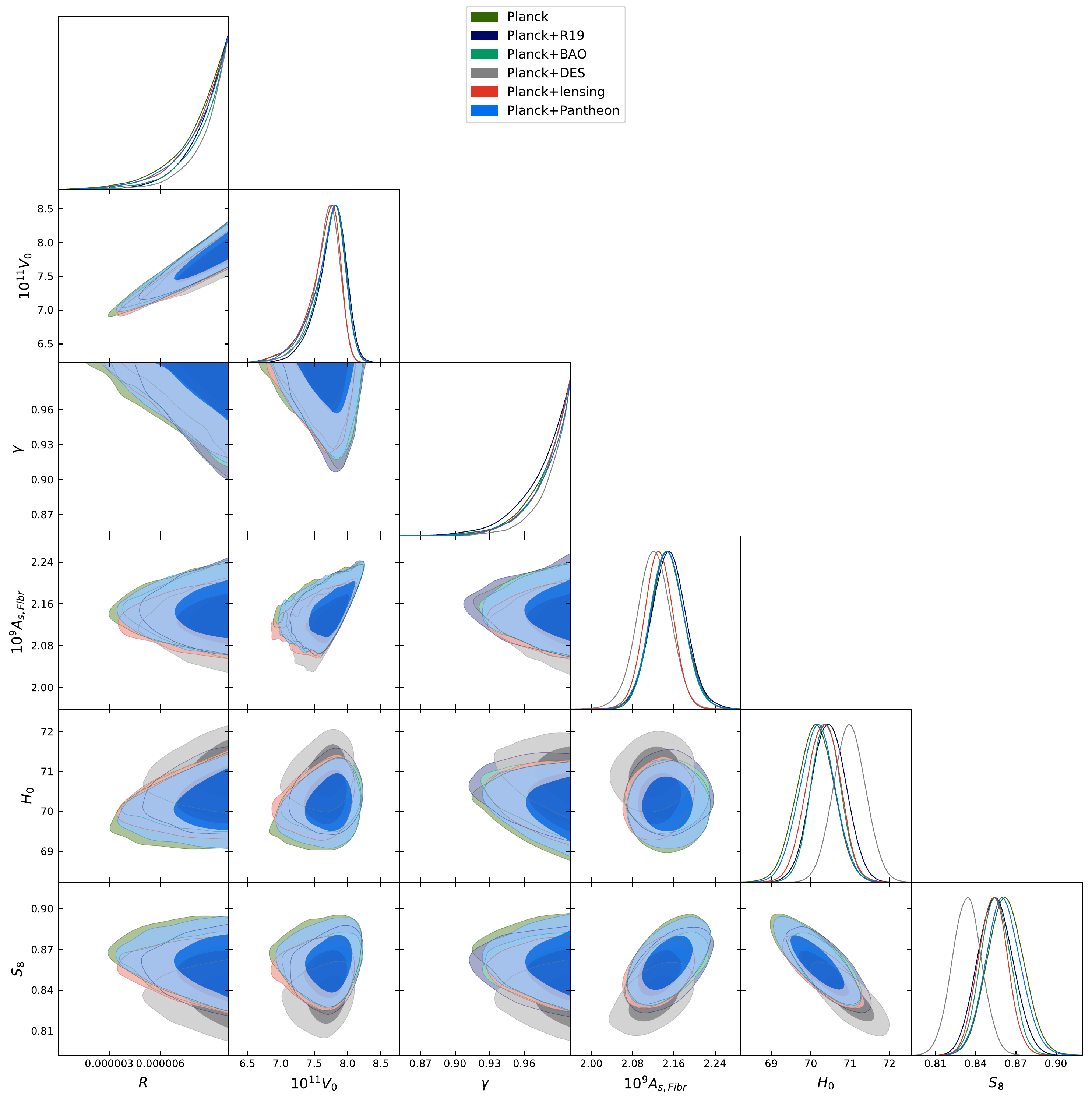}
	\caption{One dimensional posterior distributions and two-dimensional joint contours at 68\% and 95\% CL for $0 < \gamma < 1$, for the different combinations of data considered in this work.}
	\label{fig:tri_gammaUpto1}
\end{center}
\end{figure*}

In Table~\ref{tab:tri_gammaUpto1} we report the constraints at 68\% CL for the independent cosmological parameters of the FI model with $0 < \gamma < 1$ (above the horizontal line), and for some derived ones (below the horizontal line), making use of several combination of present cosmological probes. Moreover, in Fig.~\ref{fig:tri_gammaUpto1} we show a triangular plot, i.e. the 1D posterior distributions and 2D contour plots for some interesting parameters of the FI model with $0 < \gamma < 1$. 

If we compare now our results obtained for Planck alone (first column of Table~\ref{tab:tri_gammaUpto1}) with those obtained in a $\Lambda$CDM model for the same dataset, we see a shift of most of the cosmological parameters. In particular, we have that in FI models with $0 < \gamma < 1$ both $\Omega_bh^2$ and $\Omega_ch^2$ move towards larger values at many standard deviations, while $\theta$ shifts down. In our model, the amplitude and the spectral index of the primordial scalar perturbations $A_s$ and $n_s$, as well as the effective number of relativistic degrees of freedom $N_{\rm eff}$ and the tensor-to-scalar ratio $r$ are computed by using eqs. (\ref{As}), (\ref{nsrnew}), (\ref{nsr2}) and (\ref{NeffPred}) respectively, and the constraints obtained for the free parameters $\gamma$, $R$ and $V_0$. This is the reason why we have a strong prediction for $r=0.00846^{+0.00045}_{-0.00011}$ at 68\% CL different from zero, both $A_s$ and $n_s$ larger than the $\Lambda$CDM scenario, and an $N_{\rm eff}>3.046$ at many sigmas. Regarding the parameters of FI models, we find at 68\% CL and for Planck alone $\gamma>0.974$, $R>7.86\times10^{-6}$ and $V_0=(7.69^{+0.31}_{-0.15})\times 10^{-11}$. An interesting feature is the possibility of increasing the Hubble constant parameters due to the strong correlation between $N_{\rm eff}$ and $H_0$. In fact, in this scenario $H_0$ is estimated to be $H_0=70.12\pm0.47$ (km/s)/Mpc at 68\% CL, alleviating below $3$ standard deviations the very well know $4.4\sigma$ tension between the Planck~\cite{Aghanim:2018eyx} and the SH0ES~\cite{Riess:2019cxk} collaborations measurements of this parameter. Unfortunately, this scenario is disfavored by the data that show a worsening of the $\chi^2$ of $17.87$ with respect to the $\Lambda$CDM model, even if our scenario has one more degree of freedom.

If we now look at the other columns of the same Table~\ref{tab:tri_gammaUpto1}, or the contour plots in Fig.~\ref{fig:tri_gammaUpto1}, we see that our results are very robust, showing minimal shifts in the cosmological parameters by combining Planck with other independent cosmological probes. In fact, the larger shifts we can see on the cosmological parameters are for the combination Planck+DES, that however keeps almost unaltered the bounds on the cosmological parameters characteristic of FI models.

\begin{center}      
\begin{table*} 
\scalebox{0.95}{
\begin{tabular}{cccccccccccccccc}       
\hline\hline                       
Parameters & Planck   & Planck & Planck& Planck & Planck & Planck \\ 
 &  &+R19 & +BAO  & +DES & +lensing & + Pantheon \\ \hline
 
 $\Omega_b h^2$ & $    0.02260 \pm 0.00013$ &  $    0.02267\pm0.00013$ & $    0.02264\pm0.00012$ & $    0.02273\pm 0.00013$ & $    0.02263\pm 0.00013$ & $    0.02261\pm 0.00013$ \\
 
$\Omega_c h^2$ & $    0.1320^{+0.0011}_{-0.0013}$  & $    0.1313^{+0.0010}_{-0.0013}$ & $    0.1313^{+0.0009}_{-0.0010}$ & $    0.1297\pm 0.0010 $& $    0.1314^{+0.0010}_{-0.0011} $& $    0.1318^{+0.0010}_{-0.0012} $\\

$100\theta_{\rm MC}$ & $    1.03944\pm 0.00030$ &  $    1.03953\pm0.00029$ &  $    1.03951\pm 0.00029$ & $    1.03962\pm 0.00029$ & $    1.03948\pm 0.00030$& $    1.03945\pm 0.00030$\\

$\tau$ & $    0.0526\pm 0.0072$ &  $    0.0534\pm0.0075$ &  $    0.0526\pm 0.0073$ & $    0.0484\pm 0.0077$ & $    0.0493\pm0.0067$& $    0.0524\pm0.0072$\\

$\gamma$ & $    >0.974$ &  $    >0.971$ & $    >0.976$ & $    >0.980$ & $    >0.976$& $    >0.975$\\

$10^6 R$ & $    >7.86$ &  $    >8.32$ & $    >8.22$ & $    >8.41$ & $    >8.00$& $    >7.94$\\

$10^{11}V_0$ & $    7.69 ^{+0.31}_{-0.15}$ &  $    7.75^{+0.26}_{-0.16}$ & $    7.72^{+0.27}_{-0.15}$ & $    7.66 ^{+0.25}_{-0.15}$ & $    7.64^{+0.29}_{-0.13}$& $    7.70^{+0.30}_{-0.15}$\\

\hline

$H_0 $[(km/s)/Mpc] & $   70.12\pm0.47$&  $   70.46\pm 0.43$ & $   70.38\pm0.37$ & $   70.97\pm0.41$ & $   70.32\pm 0.42$& $   70.16\pm 0.45$\\

$\sigma_8$ & $    0.8406\pm0.0074$ &  $    0.8388\pm 0.0076$ & $    0.8382\pm0.0071$ & $    0.8282\pm 0.0068$ & $    0.8352\pm0.0059$& $    0.8400^{+0.0068}_{-0.0076}$\\

$S_8$ & $    0.862\pm0.014$ &  $    0.855\pm 0.013$ & $  0.855\pm0.011  $ & $    0.834\pm 0.011$ & $    0.853\pm0.011$& $    0.861\pm0.013$\\

$10^{9} A_s$ & $    2.149\pm 0.033$ &  $    2.150\pm0.034$ & $    2.146\pm 0.033$ & $    2.120\pm 0.033$ & $    2.130\pm 0.028$& $    2.148\pm 0.033$\\

$n_s$ & $    0.9724^{+0.0014}_{-0.0003}$ & $    0.9727^{+0.0011}_{-0.0003}$ &   $    0.9726^{+0.0011}_{-0.0002}$ & $    0.97275^{+0.00099}_{-0.00022}$ &  $    0.9725^{+0.0013}_{-0.0003}$&  $    0.9725^{+0.0013}_{-0.0003}$\\

$N_{\rm eff}$ & $    3.674^{+0.005}_{-0.028}$ &  $    3.678^{+0.007}_{-0.032}$ & $    3.673^{+0.004}_{-0.027}$ & $    3.667^{+0.004}_{-0.021}$ & $    3.672^{+0.005}_{-0.026}$& $    3.673^{+0.008}_{-0.027}$\\

$r$ & $    0.00846^{+0.00045}_{-0.00011}$ & $    0.00855^{+0.00036}_{-0.00009}$ &   $    0.00853^{+0.00038}_{-0.00009}$ & $    0.00857^{+0.00034}_{-0.00008}$ &  $    0.00849^{+0.00042}_{-0.00010}$&  $    0.00848^{+0.00043}_{-0.00011}$\\

\hline
$\Delta \chi_{\rm bestfit}^2$ & $    +17.87$ &  $    $ & $    $& $   $ & $    $ & $  $\\
\hline\hline                                         
\end{tabular} }
\caption{Observational constraints at 68$\%$~CL on the independent (above the line) and derived (below the line) cosmological parameters of FI models with $0 < \gamma < 1$, for the different combinations of data considered in this work. In the bottom line we quote the difference in the best-fit $\chi^2$ values with respect to the $\Lambda$CDM case for the same Planck data.}
\label{tab:tri_gammaUpto1}                          
\end{table*}                                      
\end{center}

\subsection{$\gamma = 1$}

\begin{figure*}
\begin{center}
	\includegraphics[width=0.7\linewidth]{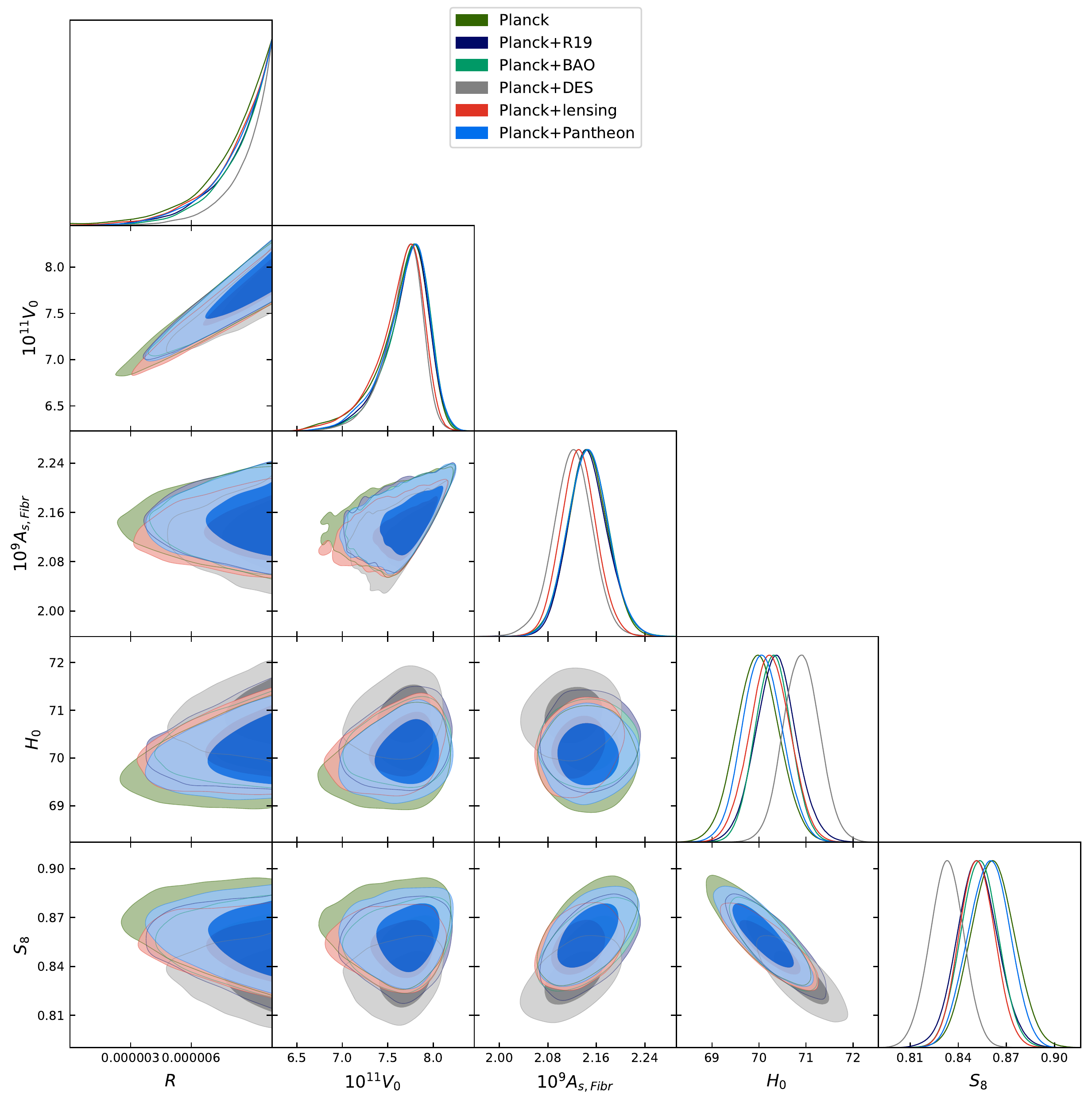}
	\caption{One dimensional posterior distributions and two-dimensional joint contours at 68\% and 95\% CL for $\gamma=1$, for the different combinations of data considered in this work.}
	\label{fig:tri_gamma1}
\end{center}
\end{figure*}

In Table~\ref{tab:tri_gamma1} we report the constraints at 68\% CL for the independent cosmological parameters of FI models with $\gamma = 1$ (above the horizontal line), and for some derived ones (below the horizontal line), making use of several combinations of present cosmological probes. Moreover, in Fig.~\ref{fig:tri_gamma1} we show a triangular plot, i.e. the 1D posterior distributions and 2D contour plots for some interesting parameters of FI models with $\gamma = 1$. 
 
Even in this $\gamma = 1$ case, if we compare our results for Planck alone dataset (first column of Table~\ref{tab:tri_gamma1}) with those obtained in a $\Lambda$CDM model, we see almost the same shift of the cosmological parameters we saw already in the $0 < \gamma < 1$ case, because $\gamma$ was consistent with $1$. For the very same reason, the constraints in FI models with $\gamma = 1$ are very similar to those obtained in the $0 < \gamma < 1$ case. In fact, when $\gamma=1$ we find at 68\% CL and for Planck alone that $R>7.70\times10^{-6}$ and $V_0=(7.66^{+0.33}_{-0.15})\times 10^{-11}$.
Regarding the prediction of the model, in this case $\gamma=1$ which is equivalent to $N_{\rm eff}=3.646$ (see eq.(\ref{NeffPred})). Moreover, we find $r=0.00842^{+0.00048}_{-0.00011}$ at 68\% CL different from zero, and again both $A_s$ and $n_s$ larger than in the $\Lambda$CDM scenario. Even in this case, and because of the strong correlation between $N_{\rm eff}$ and $H_0$, we have a large Hubble constant parameter, $H_0=69.97\pm0.46$ (km/s)/Mpc at 68\% CL, relaxing the tension below $3$ standard deviations. Unfortunately, even this scenario is disfavored by the data since they show a worsening of the $\chi^2$ of $16.41$ with respect to the $\Lambda$CDM model, for the same number of degrees of freedom.

If we now look at the other cases of the same Table~\ref{tab:tri_gamma1}, and the plots in Fig.~\ref{fig:tri_gamma1} showing Planck combined with the other cosmological probes, we see again that our results are almost unmodified, and the larger shifts are due to the combination of Planck+DES data.
 
 \begin{center}                  
\begin{table*} 
\scalebox{0.95}{
\begin{tabular}{cccccccccccccccc}  
\hline\hline                        
Parameters & Planck   & Planck & Planck& Planck & Planck & Planck \\ 
 &  &+R19 & +BAO  & +DES & +lensing & + Pantheon \\ \hline
 
 $\Omega_b h^2$ & $    0.02259 \pm 0.00013$ &  $    0.02266\pm0.00013$ & $    0.02264\pm0.00013$ & $    0.02274\pm 0.00013$ & $    0.02262\pm 0.00013$ & $    0.02261\pm 0.00013$ \\
 
$\Omega_c h^2$ & $    0.1315\pm0.0011$  & $    0.1306\pm0.0010$ & $    0.13073\pm0.00087$ & $    0.12928\pm 0.00088 $& $    0.13085\pm0.00098 $& $    0.1312\pm0.0010 $\\

$100\theta_{\rm MC}$ & $    1.03950\pm 0.00029$ &  $    1.03960\pm0.00029$ &  $    1.03959\pm 0.00028$ & $    1.03967\pm 0.00028$ & $    1.03955\pm 0.00030$& $    1.03952\pm 0.00029$\\

$\tau$ & $    0.0522\pm 0.0074$ &  $    0.0531\pm0.0073$ &  $    0.0530\pm 0.0075$ & $    0.0490\pm 0.0075$ & $    0.0498\pm0.0069$& $    0.0526\pm0.0073$\\

$10^6 R$ & $    >7.70$ &  $    >7.99$ & $    >8.08$ & $    >8.47$ & $    >7.87$& $    >7.93$\\

$10^{11}V_0$ & $    7.66 ^{+0.33}_{-0.15}$ &  $    7.70^{+0.29}_{-0.16}$ & $    7.71^{+0.28}_{-0.15}$ & $    7.66 ^{+0.25}_{-0.14}$ & $    7.62^{+0.31}_{-0.15}$& $    7.69^{+0.30}_{-0.17}$\\

\hline

$H_0 $[(km/s)/Mpc] & $   69.97\pm0.46$&  $   70.36\pm 0.43$ & $   70.30\pm0.38$ & $   70.90\pm0.39$ & $   70.23\pm 0.43$& $   70.08\pm 0.42$\\

$\sigma_8$ & $    0.8394\pm0.0074$ &  $    0.8368\pm 0.0073$ & $    0.8372\pm0.0073$ & $    0.8276\pm 0.0067$ & $    0.8346\pm0.0060$& $    0.8388\pm0.0073$\\

$S_8$ & $    0.862\pm0.014$ &  $    0.852\pm 0.013$ & $  0.853\pm0.012  $ & $    0.833\pm 0.011$ & $    0.852\pm0.011$& $    0.859\pm0.013$\\

$10^{9} A_s$ & $    2.146\pm 0.033$ &  $    2.146^{+0.030}_{-0.034}$ & $    2.146\pm 0.034$ & $    2.121\pm 0.033$ & $    2.131\pm 0.030$& $    2.146\pm 0.033$\\

$n_s$ & $    0.9723^{+0.0015}_{-0.0003}$ & $    0.9725^{+0.0013}_{-0.0003}$ &   $    0.9726^{+0.0012}_{-0.0003}$ & $    0.97277^{+0.00098}_{-0.00019}$ &  $    0.9724^{+0.0014}_{-0.0003}$&  $    0.9725^{+0.0013}_{-0.0003}$\\

$r$ & $    0.00842^{+0.00048}_{-0.00011}$ & $    0.00849^{+0.00042}_{-0.00011}$ &   $    0.00851^{+0.00040}_{-0.00010}$ & $    0.00858^{+0.00033}_{-0.00007}$ &  $    0.00846^{+0.00045}_{-0.00011}$&  $    0.00848^{+0.00043}_{-0.00011}$\\
\hline
$\Delta \chi_{\rm bestfit}^2$ & $    +16.41 $ &  $   $ & $    $& $   $ & $   $ & $ $\\
\hline\hline                         
\end{tabular} }
\caption{Observational constraints at 68$\%$~CL on the independent (above the line) and derived (below the line) cosmological parameters of FI models with $\gamma = 1$, for the different combinations of data considered in this work. In the bottom line we quote the difference in the best-fit $\chi^2$ values with respect to the $\Lambda$CDM case for the same Planck data.}
\label{tab:tri_gamma1}                      
\end{table*}                         
\end{center}

\subsection{$1 < \gamma \leq 20$}

\begin{figure*}
\begin{center}
	\includegraphics[width=0.7\linewidth]{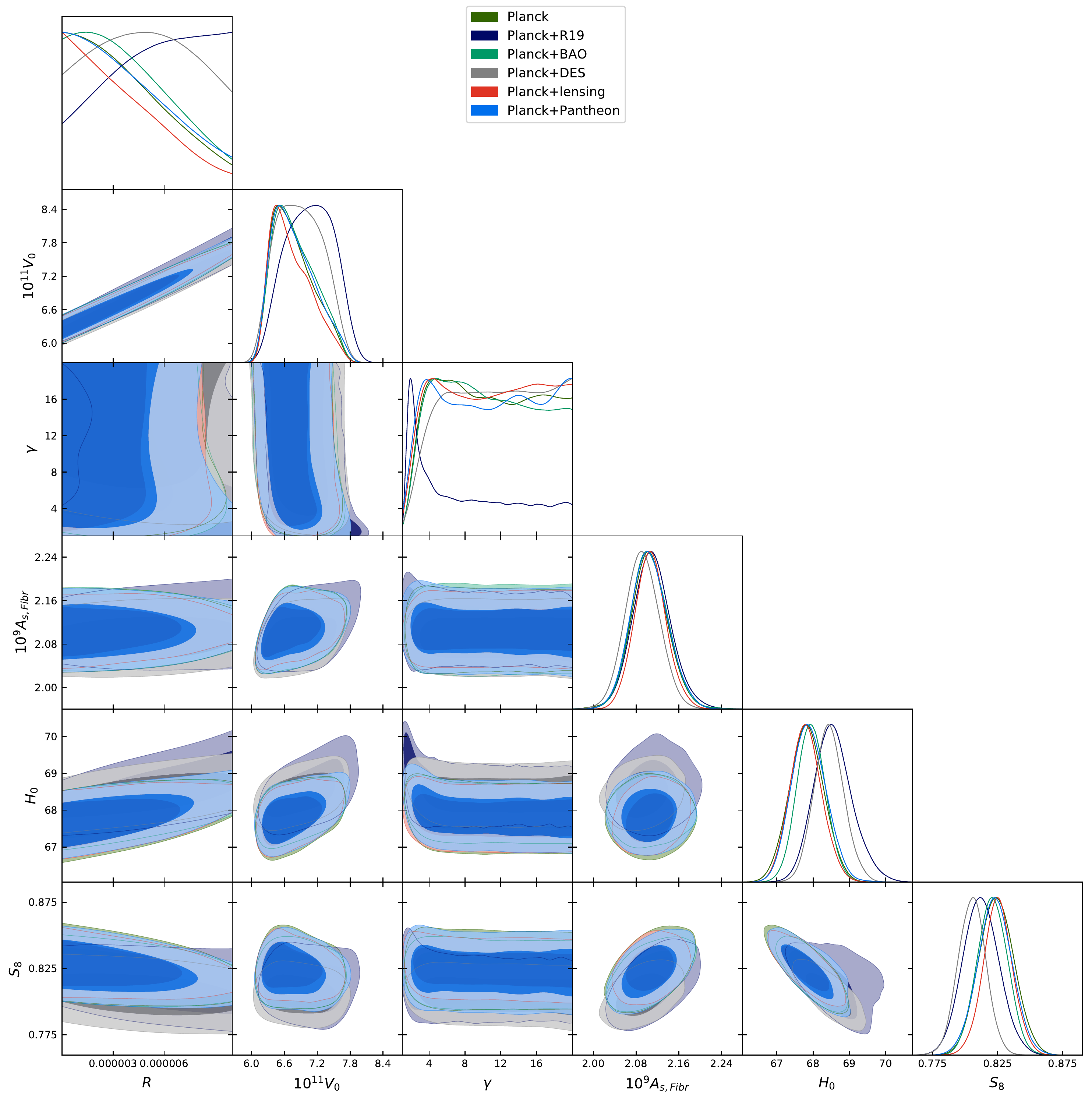}
	\caption{One dimensional posterior distributions and two-dimensional joint contours at 68\% and 95\% CL for $1 < \gamma \leq 20$, for the different combinations of data considered in this work.}
	\label{fig:tri_gammaUpto20}
\end{center}
\end{figure*}

Finally, in Table~\ref{tab:tri_gammaUpto20} we report the bounds at 68\% CL for the independent cosmological parameters of FI models with $1 < \gamma \leq 20$ (above the horizontal line), and for some derived ones (below the horizontal line), combining different present cosmological probes. Moreover, in Fig.~\ref{fig:tri_gammaUpto20} we show a triangular plot, comprising 1D posterior distributions and 2D contour plots for some interesting parameters of FI models with $1 < \gamma \leq 20$. 

In this $1 < \gamma \leq 20$ case, if we compare our results obtained for Planck alone (first column of Table~\ref{tab:tri_gammaUpto20}) with those obtained in a $\Lambda$CDM model for the same dataset, we see that most of the cosmological parameters are perfectly in agreement within a standard deviation, on the contrary of the previous $\gamma$ cases. Moreover, in this FI scenario with $1 < \gamma \leq 20$, we find at 68\% CL for Planck alone $\gamma>7.41$, and therefore $N_{\rm eff}=3.062^{+0.004}_{-0.015}$ at 68\% CL is here in agreement with its expected value $3.046$~\cite{Mangano:2005cc,deSalas:2016ztq}. In this case, instead of having a lower limit for $R$ as in the previous $\gamma$ cases, we find an upper limit $R<4.80\times10^{-6}$ at 68\% CL, while $V_0$ shifts too towards lower values $V_0=(6.76^{+0.25}_{-0.49})\times 10^{-11}$ at 68\% CL because of the positive correlation between these two parameters, as we can see in Fig.~\ref{fig:tri_gammaUpto20}.
Even in this FI scenario with $1 < \gamma \leq 20$ we have a strong prediction for $r=0.00731^{+0.00026}_{-0.00072}$ at 68\% CL different from zero, while both $A_s$ and $n_s$ are consistent with the constraints obtained in a $\Lambda$CDM scenario.
The possibility of increasing the Hubble constant parameters because of the correlation between $N_{\rm eff}$ and $H_0$ is lost in this case, because $N_{\rm eff}$ is in agreement with the standard expectation. In fact, here we have $H_0=67.82\pm0.47$ (km/s)/Mpc at 68\% CL, shifted one sigma towards a larger value, but still at $4.1\sigma$ tension with the R19 estimate~\cite{Riess:2019cxk}. FI models with $1 < \gamma \leq 20$ are the most favored by the data between those explored in this work, worsening the $\chi^2$ of $0.39$ with respect to the $\Lambda$CDM model, having just one more degree of freedom.

If we now look at the other cases of the same Table~\ref{tab:tri_gammaUpto20}, and the plots in Fig.~\ref{fig:tri_gammaUpto20} showing Planck combined with the other cosmological probes, we see again that our results are more dataset dependent than the other $\gamma$ cases we explored before. The only combinations that give similar results to the Planck alone case are Planck+lensing and Planck+Pantheon, while the other ones deserve a more complete discussion.

In particular, in the Planck+R19 case, i.e. the third column of Table~\ref{tab:tri_gammaUpto20}, we are adding to the Planck data a gaussian prior on $H_0$ similar to R19, and this prior is at $4.1\sigma$ tension with the Hubble constant estimated by Planck alone. Hence we are combining datasets in disagreement and the results are not completely reliable. For this very same reason, we see one sigma shift of $H_0$ in the R19 direction, and due to a positive correlation with this parameter, we have also a one sigma shift of $V_0$ towards higher values, i.e. $V_0=(7.06^{+0.47}_{-0.43})\times 10^{-11}$ at 68\% CL, and $N_{\rm eff}$ larger than the expected value at more that one standard deviation. Consequently we obtain $\gamma=8.4^{+11}_{-7.2}$ at 68\% CL, instead of just a lower limit. Moreover, due to the strong $V_0$-$R$ positive correlation, we see that instead of an upper limit like in the Planck alone case, we have now a lower limit for $R>4.13\times10^{-6}$ at 68\% CL.

If we now look at the Planck+BAO combination, i.e. the fourth column of the same Table~\ref{tab:tri_gammaUpto20}, we see that the constraints on the cosmological parameters are very similar to the ones obtained in the Planck alone case, with the exception of $\gamma$ that is now fully bounded at 68\% CL, i.e. $\gamma=10.7^{+4.2}_{-7.5}$. However, looking at the 1D posterior distribution of $\gamma$ in Fig.~\ref{fig:tri_gammaUpto20}, we see that in the Planck+BAO combination there is no actual peak that can justify this bound.

Finally, in the Planck+DES case, as in the other $\gamma$ scenarios, $S_8$ is shifted towards lower values, more in agreement with the cosmic shear findings, and consequently $H_0$ moves towards slightly larger values due to their negative correlation, as we can see in Fig.~\ref{fig:tri_gammaUpto20}.

\begin{center}                              
\begin{table*} 
\scalebox{0.95}{
\begin{tabular}{cccccccccccccccc}       
\hline\hline                                        
Parameters & Planck   & Planck & Planck& Planck & Planck & Planck \\ 
 &  &+R19 & +BAO  & +DES & +lensing & + Pantheon \\ \hline
 
 $\Omega_b h^2$ & $    0.02244 \pm 0.00014$ &  $    0.02256\pm0.00014$ & $    0.02246\pm0.00013$ & $    0.02254\pm 0.00014$ & $    0.02244\pm 0.00014$ & $    0.02245\pm 0.00014$ \\
 
$\Omega_c h^2$ & $    0.1194^{+0.0010}_{-0.0012}$  & $    0.1189^{+0.0009}_{-0.0022}$ & $    0.1191^{+0.0009}_{-0.0010}$ & $    0.11794\pm 0.00095 $& $    0.1195^{+0.0010}_{-0.0011} $& $    0.1193^{+0.0010}_{-0.0012} $\\

$100\theta_{\rm MC}$ & $    1.04098\pm 0.00030$ &  $    1.04105^{+0.00040}_{-0.00032}$ &  $    1.04101\pm 0.00029$ & $    1.04111\pm 0.00029$ & $    1.04096\pm 0.00030$& $    1.04098^{+0.00032}_{-0.00028}$\\

$\tau$ & $    0.0563\pm 0.0079$ &  $    0.0576^{+0.0071}_{-0.0085}$ &  $    0.0567^{+0.0070}_{-0.0082}$ & $    0.0549\pm 0.0075$ & $    0.0565\pm0.0072$& $    0.0561\pm0.0078$\\

$\gamma$ & $    >7.41$ &  $    8.4^{+11}_{-7.2}$ & $    10.7^{+4.2}_{-7.5}$ & $    >8.31$ & $    >7.56$& $    >7.37$\\

$10^6 R$ & $    <4.80$ &  $    >4.13$ & $    <5.07$ & unconstr. & $    <4.51$& $    <5.02$\\

$10^{11}V_0$ & $    6.76 ^{+0.25}_{-0.49}$ &  $    7.06^{+0.47}_{-0.43}$ & $    6.79^{+0.28}_{-0.49}$ & $    6.88 ^{+0.39}_{-0.48}$ & $    6.73^{+0.24}_{-0.47}$& $    6.78^{+0.27}_{-0.50}$\\

\hline

$H_0 $[(km/s)/Mpc] & $   67.82\pm0.47$&  $   68.56^{+0.46}_{-0.57}$ & $   67.95\pm0.40$ & $   68.44\pm0.41$ & $   67.80^{+0.41}_{-0.46}$& $   67.89^{+0.44}_{-0.50}$\\

$\sigma_8$ & $    0.8110\pm0.0076$ &  $    0.8090^{+0.0079}_{-0.0089}$ & $    0.8102\pm0.0073$ & $    0.8034\pm 0.0065$ & $    0.8115\pm0.0061$& $    0.8107\pm0.0076$\\

$S_8$ & $    0.824\pm0.013$ &  $    0.812\pm 0.014$ & $  0.821\pm0.012  $ & $    0.805\pm 0.011$ & $    0.825\pm0.011$& $    0.823\pm0.013$\\

$10^{9} A_s$ & $    2.105\pm 0.035$ &  $    2.110^{+0.033}_{-0.037}$ & $    2.106^{+0.032}_{-0.036}$ & $    2.092\pm 0.032$ & $    2.107\pm 0.030$& $    2.105^{+0.032}_{-0.036}$\\

$n_s$ & $    0.9696^{+0.0010}_{-0.0026}$ & $    0.9709^{+0.0028}_{-0.0015}$ &   $    0.9698^{+0.0012}_{-0.0026}$ & $    0.9705^{+0.0019}_{-0.0026}$ &  $    0.9695^{+0.0010}_{-0.0025}$&  $    0.9697^{+0.0011}_{-0.0026}$\\

$N_{\rm eff}$ & $    3.062^{+0.004}_{-0.015}$ &  $    3.098^{+0.004}_{-0.054}$ & $    3.062^{+0.004}_{-0.015}$ & $    3.059^{+0.003}_{-0.012}$ & $    3.063^{+0.005}_{-0.016}$& $    3.064^{+0.006}_{-0.018}$\\

$r$ & $    0.00731^{+0.00026}_{-0.00072}$ & $    0.00774^{+0.00076}_{-0.00060}$ &   $    0.00735^{+0.00030}_{-0.00073}$ & $    0.00756^{+0.00046}_{-0.00083}$ &  $    0.00726^{+0.00025}_{-0.00067}$&  $    0.00734^{+0.00029}_{-0.00074}$\\

\hline
$\Delta \chi_{\rm bestfit}^2$ & $    +0.39 $ &  $   $ & $   $& $   $ & $    $ & $  $\\
\hline\hline                                   
\end{tabular} }
\caption{Observational constraints at 68$\%$~CL on the independent (above the line) and derived (below the line) cosmological parameters of FI models with $1 < \gamma \leq 20$, for the different combinations of data considered in this work. In the bottom line we quote the difference in the best-fit $\chi^2$ values with respect to the $\Lambda$CDM case for the same Planck data.}
\label{tab:tri_gammaUpto20}                                          
\end{table*}                                    
\end{center}

\section{Bounds on microscopic parameters}
\label{MicroscopicBounds}

In this section we shall follow the notation of \cite{Cicoli:2018cgu} and translate the bounds on $\gamma$, $R$ and $V_0$ into bounds on the microscopic parameters of the FI landscape which depend on discrete 3-form bulk fluxes and 2-form brane fluxes.  

FI models are characterised by the following microscopic parameters: 
\begin{itemize}
\item $2$ topological properties which depend on the choice of the underlying Calabi-Yau manifold: the Calabi-Yau Euler number $\xi$ which is expected to be of order unity and the intersection number $k_{122}$ which takes in general $\mathcal{O}(1-10)$ values, and so we shall set $k_{122}=5$.

\item $1$ discrete quantity which depends on the number $N$ of D$7$-branes wrapped around the blow-up cycle which supports non-perturbative effects. $N$ is constrained by tadpole cancellation, and in general one obtains $\mathcal{O}(1-10)$ values. 

\item $6$ discrete parameters which are functions of either the dilaton or complex structure moduli, and so depend on the choice of 3-form background fluxes: the string coupling $g_s$, the value of the tree-level superpotential $W_0$, the prefactor of non-perturbative effects $A_3$ and $3$ coefficients of string loop corrections to the K\"ahler potential $c_1^{\rm KK}$, $c_2^{\rm KK}$ and $c^{\rm W}$. Natural values of these last $3$ quantities are expected to be in the range between $0.1$ and $10$. In what follows we shall therefore set $c_1^{\rm KK} = c^{\rm W}=4$ and $c_2^{\rm KK}=0.1$. 

\item $1$ discrete quantity $n_2$ which determines the quantised 2-form gauge flux on the worldvolume of the D$7$-branes wrapped around the fibre divisor which realise the visible sector.
\end{itemize}

Notice that the $3$ microscopic parameters $\xi$, $N$ and $A_3$ enter only in the determination of the extra-dimensional volume $\mathcal{V}$. Hence in our analysis, we shall trade $\xi$, $A_3$ and $N$ for the single parameter $\mathcal{V}$. 

The $3$ underlying parameters $\gamma$, $R$ and $V_0$ are functions of the $4$ microscopic parameters which we left over as free: $\gamma=\gamma(g_s, W_0,\mathcal{V}, n_2)$, $R=R(g_s)$, $V_0=V_0(g_s, W_0, \mathcal{V})$. We can therefore constrain $g_s$, $W_0$, $\mathcal{V}$ and $n_2$ by using the observational constraints on $\gamma$, $R$ and $V_0$ obtained in Sec. \ref{Results}, supplemented with the phenomenological requirement $\alpha_{\rm vis} = \alpha_{\rm vis}(g_s, W_0, \mathcal{V}, n_2) =1/25$.

Moreover, notice that $\gamma$ can be written as in (\ref{tau1Rel}) where $\tau_1$ depends on $g_s$ and $\mathcal{V}$. Hence we can consider $\alpha_{\rm vis} = \alpha_{\rm vis}(g_s, W_0, \mathcal{V}, n_2) =1/25$ as a constraint which gives $n_2$ once $g_s$, $W_0$ and $\mathcal{V}$ have already been bounded by using the results of Sec. \ref{Results} with $\gamma=\gamma(g_s, \mathcal{V})$, $R=R(g_s)$ and  $V_0=V_0(g_s, W_0, \mathcal{V})$. 

\begin{figure}[!htbp]
\centering
\includegraphics[scale = 0.4]{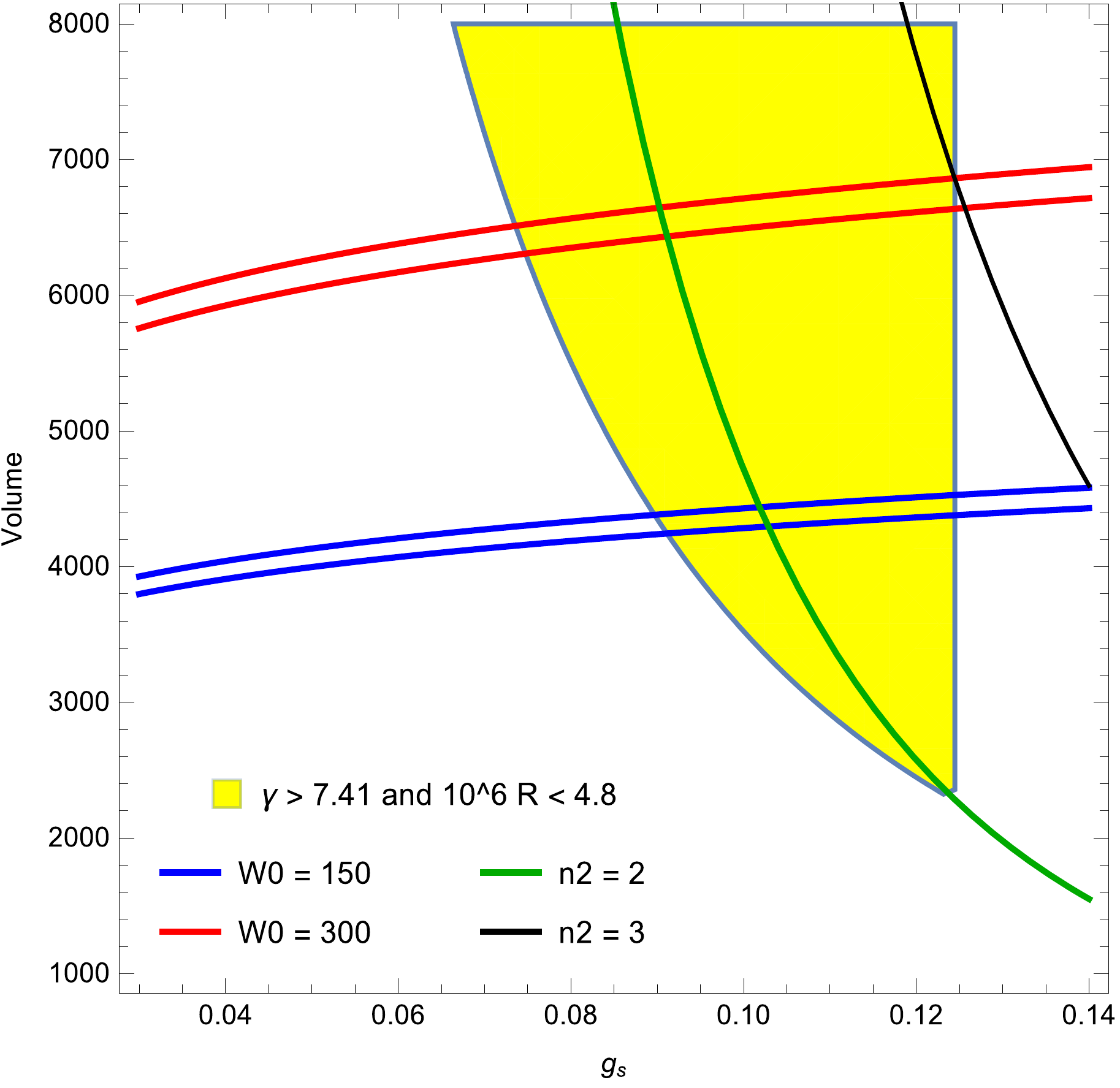}
\caption{Phenomenological bounds on the microscopic parameters $\mathcal{V}$ and $g_s$ for $\gamma$ in $[1,20]$ and Planck data alone. The yellow region corresponds to $\gamma> 7.41$ and $R< 4.8\times 10^{-6}$. The two blue and red lines correspond to the upper and lower bounds on $V_0$ at $68\%$ CL for $W_0=150$ and $W_0=300$. The green and black lines correspond to $\alpha_{\rm vis}^{-1}=25$ for $n_2=2$ and $n_2=3$.}
\label{Fig8}
\end{figure}

We shall focus on the case with $1<\gamma\leq 20$ since this is the one which is statistically favoured by observations, and consider Planck data alone. Our results are displaced in Fig. \ref{Fig8} for the $(\mathcal{V}, g_s)$-parameter space. The yellow region corresponds to $\gamma> 7.41$ and $R< 4.8\times 10^{-6}$, the two blue and red lines correspond to the upper and lower bounds on $V_0$ at $68\%$ CL for $W_0=150$ and $W_0=300$ respectively, whereas the green and black lines correspond to $\alpha_{\rm vis}^{-1}=25$ for $n_2=2$ and $n_2=3$ respectively. 

Notice that the comparison with cosmological observations constrains the discrete gauge flux parameter $n_2$ very well since the curve corresponding to $\alpha_{\rm vis}^{-1}=25$ intersects the yellow region only for $n_2=2$ and $n_2=3$. Given that $W_0$ is upper bounded by tadpole cancellation which gives maximal value of order $500$, we also realise that the string coupling is forced to lie in the range $0.065\lesssim g_s\lesssim 0.125$ and the Calabi-Yau volume $2500\lesssim\mathcal{V}\lesssim 9000$.

\section{Conclusions}
\label{Conclusions}

In this paper we showed that predictions from string theory can indeed be put to the experimental test. In particular we focused on Fibre Inflation which is a class of type IIB string inflationary models that feature an underlying landscape of microscopic flux-dependent parameters. Similar studies in string cosmology have also been performed in \cite{Bhattacharya:2017pws, Bhattacharya:2020gnk} using however a different methodology.

FI models have been studied in great detail, determining not just the inflationary dynamics but also the post-inflationary evolution including reheating and the potential production of extra neutrino-like species. Thanks to this detailed analysis, these models are ready to be confronted with observations. 

In our analysis we included several recent cosmological data coming from Planck, direct measurements of $H_0$, BAO, DES, CMB lensing and Pantheon. We focused on a $7$-dimensional baseline space described by the standard parameters $\Omega_bh^2$, $\Omega_{c}h^2$, $100 \theta_{MC}$ and $\tau$, with in addition $3$ combinations of microscopic parameters $\gamma$, $R$ and $V_0$ which characterise Fibre Inflation. After imposing flat priors on each of these parameters, we derived bounds on $A_s$, $n_s$, $r$ and $N_{\rm eff}$ for different ranges of $\gamma$.

We found that the range of $\gamma$ which gives the best fit to these recent cosmological data is $1<\gamma\leq 20$. In particular, for Planck data alone we find at $68\%$ CL $\gamma>7.41$, $R<4.80\times 10^{-6}$ and $10^{11}\, V_0=6.76^{+0.25}_{-0.49}$ together with $n_s=0.9696^{+0.0010}_{-0.0026}$, $N_{\rm eff}=3.062^{+0.004}_{-0.015}$ and $r=0.00731^{+0.00026}_{-0.00072}$. The prediction for the tensor-to-scalar ratio is particularly promising since it might be tested by the next generation of cosmological observations.

From the microscopical point of view, this implies that the models in the Fibre Inflation landscape which are statistically favoured by cosmological data are the ones leading to the case \textit{small extra dark radiation} of Sec.~\ref{SecB}. In this case horizon exit takes place in the plateau far away from the exponential steepening of the potential, so leading to no power loss at large angular scales. Moreover $N_{\rm eff}$ is very close to the Standard Model values, implying that the inflaton decay into bulk ultra-light axions has to be suppressed by the presence of a non-zero gauge on the D7-brane stack wrapped around the fibre divisor which realises the visible sector. 

In Sec.~\ref{MicroscopicBounds} we finally translated the previous bounds into constraints on the microscopic flux dependent quantities, showing how agreement with cosmological observations forces the string coupling to lie in the range $0.065\lesssim g_s\lesssim 0.125$ and the Calabi-Yau volume in $2500\lesssim\mathcal{V}\lesssim 9000$. 

Let us stress again that this analysis illustrates how large portions of the string landscape can be ruled out by comparison with observations, in particular thanks to the presence in string models of correlations between different theoretical and phenomenological features. A crucial correlation, which could constrain further the parameter space of these 4D string models and which we did not take into account in this analysis, is the connection with particle physics predictions like those concerning supersymmetry and the QCD axion. We leave this investigation for future work. 

Let us finally mention that several `swampland conjectures' have been recently proposed based on different quantum gravity arguments \cite{Palti:2019pca}. According to these conjectures, models of inflation from string theory lack control over the effective field theory, and so are incompatible with quantum gravity. However at the moment this issue is far from being settled and it is the subject of a lively debate. A recent critical discussion of progress and open issues in controlling perturbative and non-perturbative corrections in string compactifications can for example be found in \cite{Cicoli:2018kdo}.

Focusing in particular on Fibre Inflation, these models have been shown to be embeddable in Calabi-Yau compactifications built as hypersurfaces in toric varieties with an explicit orientifold involution and a chiral brane setup which satisfies global consistency requirements like tadpole cancellation \cite{Cicoli:2011it, Cicoli:2016xae, Cicoli:2017axo}. These compactifications have also all the required higher-dimensional features to give rise to the desired corrections to the low-energy effective action which stabilise the moduli and generate the inflation potential given in (\ref{Inflationpot}). Moreover so far no quantum correction to the inflationary potential has been found which could destroy Fibre Inflation. Hence these models seem to be counter-examples to swampland conjectures. However in order to provide a final answer to this crucial open issue, one should be able to perform a systematic analysis of all possible $\alpha'$ and $g_s$ corrections to the 4D effective action which is at present a hard technical problem. A step forward towards achieving this goal has been recently performed in \cite{Burgess:2020qsc} where the authors tried to classify all possible perturbative corrections to the effective action of string compactification by using approximate symmetries like supersymmetry, scale invariance and shift symmetries.

\section*{Acknowledgments}

EDV acknowledges support from the European Research Council in the form of a Consolidator Grant with number 681431.

\appendix

\section{Upper bound on $\gamma$}
\label{AppA}

In this appendix we shall again follow the notation of \cite{Cicoli:2018cgu} and estimate an upper bound on $\gamma$ based on the consistency of the underlying UV theory. The parameter $\gamma$ looks like:
\be
\gamma = 2\alpha_{\rm vis}\langle \tau_1\rangle
\ee
where $\alpha_{\rm vis} = g^2/(4\pi)$ gives the visible sector gauge coupling while $\langle \tau_1\rangle$ is the value at the minimum of the K\"ahler modulus whose real part controls the volume of the K3 or T$^4$ divisor. This modulus is stabilised at:
\be
\langle \tau_1\rangle = g_s^{4/3} \lambda \mathcal{V}^{2/3}
\label{tau1Rel}
\ee
where $g_s$ is the string coupling, $\mathcal{V}$ is the Calabi-Yau volume in string units and $\lambda$ is expressed in terms of the coefficients of string loop corrections to the K\"ahler potential as (setting $k_{122}=5$):
\be
\lambda = 2 \cdot 5^{1/3}\,\frac{\left(c_1^{\rm KK}\right)^{4/3}}{\left(c^{\rm W}\right)^{2/3}}
\ee
Hence $\gamma$ becomes:
\be
\gamma =  4 \cdot 5^{1/3}\,\alpha_{\rm vis} \left(g_s\,c_1^{\rm KK}\right)^{4/3}\left(\frac{\mathcal{V}}{c^{\rm W}}\right)^{2/3}
\label{gammaexpression}
\ee
We now impose the following phenomenological and theoretical consistency constraints: 
\begin{enumerate}
\item A realistic GUT-like value of the gauge coupling: $\alpha_{\rm vis}^{-1}= 25$
\item Effective field theory in the perturbative regime: $g_s\lesssim 0.125$
\item Correct amplitude of the density perturbations: $\mathcal{V}\lesssim 10^4$
\item Natural values of the coefficients of the string loop corrections: $c_1^{\rm KK}=c^{\rm W}= 4$
\end{enumerate}
Applying these constraints to (\ref{gammaexpression}), we end up with the following upper bound on $\gamma$:
\be
\gamma \lesssim 20\,.
\ee

\end{document}